\documentclass[prb,preprint]{revtex4}

\pdfoutput=1

\usepackage{graphicx}
\usepackage{latexsym}
\usepackage{amssymb}
\usepackage{fourier}

\begin{document}


\title{A simple model of the strange-metal phase in cuprates}

\medskip 

\date{October 12, 2021} \bigskip

\author{Manfred Bucher \\}
\affiliation{\text{\textnormal{Physics Department, California State University,}} \textnormal{Fresno,}
\textnormal{Fresno, California 93740-8031} \\}

\begin{abstract}
Hole doping of superconducting cuprates generates lattice defects of $O$ atoms.
At and beyond closing of the pseudogap, $p \ge p^*$, they form a highly symmetric superlattice.
Umklapp processes involving reciprocal lattice vectors associated with both the host lattice and the $O$ superlattice could account for the linear temperature dependence of resistivity.

\end{abstract}

\maketitle

Leaving semiconductors and semimetals aside, classical condensed-matter physics distinguishes materials by their electric conductivity as metals or insulators. Quantum mechanics explains the distinction by the materials' bandstructure in terms of continuous (unfilled) higher-energy quantum states in the conduction band(s) of a metal but a bandgap between the (filled) valence and (unfilled) conduction band of an insulator. The quantum states of the relevant electrons are delocalized as waves in metals but localized as atomic or molecular orbitals in insulators. It came therefore as a surprise when in hole-doped cuprate high-$T_c$ superconductors a \textit{pseudogap} phase was observed, with most electrons localized but some \textit{de}localized—a mixture of insulator and metal. In that pseudogap phase the delocalized quantum states form pieces of a (here, two-dimensional) Fermi surface in reciprocal space, called \textit{Fermi arcs}, 
$\widearc{F} ={\widearc{\hat{Q}\dot{Q}\hat{Q}}}$, each extending from a ``Fermi dot'' 
$\dot{Q} = (\pm\dot{q},\pm\dot{q})$---also called ``node''---bilaterally to Fermi-arc tips $\hat{Q}$, with $\dot{q} = 0.21 \pm 0.005$ r.l.u. (in reciprocal lattice units). In the underdoped range the Fermi arcs are centered at the closest $M = (\pm\frac{1}{2},\pm\frac{1}{2})$ points of the Brillouin zone. Such a Fermi surface is called ``hole-like,'' in contrast to an ``electron-like'' Fermi surface, centered at the origin of the Brillouin-zone, $\Gamma = (0,0)$, as in highly overdoped samples.
The length of each Fermi arc increases (roughly proportionally) with  hole-doping $p$ of the host crystal, $\widearc{F} \sim p$. 
At the doping $p^*$ where the pseudogap closes (at $T = 0$, quantum critical point), the Fermi arcs join to form a \textit{complete} Fermi surface, coincident with 
a transition of the pseudogap phase to a Fermi-liquid metal. Here another surprise awaits.

In a \textit{conventional} Fermi liquid, the electron-electron scattering rate 
is given as
\begin{equation}
  \frac{1}{\tau} \approx \frac{\epsilon_F}{\hbar}\left({\frac{k_B T}{\epsilon_F}}\right)^2 ,
\end{equation}
with Fermi energy $\epsilon_F$, reduced Planck constant $\hbar = h/2\pi$, and Boltzmann constant $k_B$.\cite{1,2} 
Inserted into the Drude formula of resistivity, this gives the contribution to a conventional Fermi-liquid metal, proportional to the square of the temperature,
\begin{equation}
  \rho_{e,e} = \frac{m^*}{ne^2\tau} \approx  \frac{m^*}{ne^2}\frac{\epsilon_F}{\hbar}\left({\frac{k_BT}{\epsilon_F}}\right)^2,
\end{equation}
with effective electron mass $m^*$, electron charge $e$, and electron density $n$.
Involved in the derivation of Eqs. (1, 2) are \textit{umklapp} processes---backfolding of scattered quantum states by a reciprocal lattice vector $\mathbf{q_0}$---in $(e,e)$ scattering from quantum states near the Fermi surface.\cite{2}

The new surprise is that in the Fermi-liquid phase of high-$T_c$ cuprates, $p > p^*$, a \textit{linear} temperature dependence of the resistivity is observed, $\rho_{e,e} \propto T$,  in contrast to the $T^2$-dependence of Eq. (2). Independently, it has been noticed that the pseudogap-phase $\rightarrow$ Fermi-liquid transition in those compounds occurs when lattice-defect $O$ atoms, generated by doped holes, form a highly symmetric superlattice.\cite{3} Thus, one can distinguish \textit{two} sets of Brillouin zones in reciprocal space: one due to the reciprocal host lattice and one due to the reciprocal defect-superlattice, with corresponding reciprocal lattice vectors $\mathbf{q_0}$ and $\mathbf{Q_0}$, respectively.

We want to assume that the \textit{fraction} of the Fermi surface that is affected by one umklapp or the other depends on the \textit{relative strength} of the host-lattice and superlattice, $1-\sigma$ and $\sigma$, respectively.
As a \textit{trial} we expand Eq. (1) as
\begin{equation}
  \frac{1}{\tau} \approx \frac{\epsilon_F}{\hbar}\left[(1-\sigma)\left({\frac{k_B T}{\epsilon_F}}\right)^{2(1-\sigma)} + \sigma\left({\frac{k_B T}{\epsilon_F}}\right)^{2\sigma}\right] .
\end{equation}
Cases of Eq. (3) for $\sigma = 0, 0.1, ..., 0.5$ are graphed in Fig. 1. In the limiting case of only one kind of Brillouin zone, $\sigma = 0$, Eq. (3) reduces to Eq. (1). If two sets of  Brillouin zones are present, associated with reciprocal host lattice and defect-superlattice of \textit{equal} strength, $\sigma = \frac{1}{2}$, then Eq. (3) gives a linear temperature dependence of the scattering rate, called the ``Planckian limit,''\cite{4}
\begin{equation}
  \frac{1}{\tau} \approx \frac{\epsilon_F}{\hbar}\left[\frac{1}{2}\left({\frac{k_B T}{\epsilon_F}}\right)^{2\frac{1}{2}} + \frac{1}{2}\left({\frac{k_B T}{\epsilon_F}}\right)^{2\frac{1}{2}}\right] = \frac{k_B T}{\hbar}\;.
\end{equation}

\includegraphics[width=5.6in]{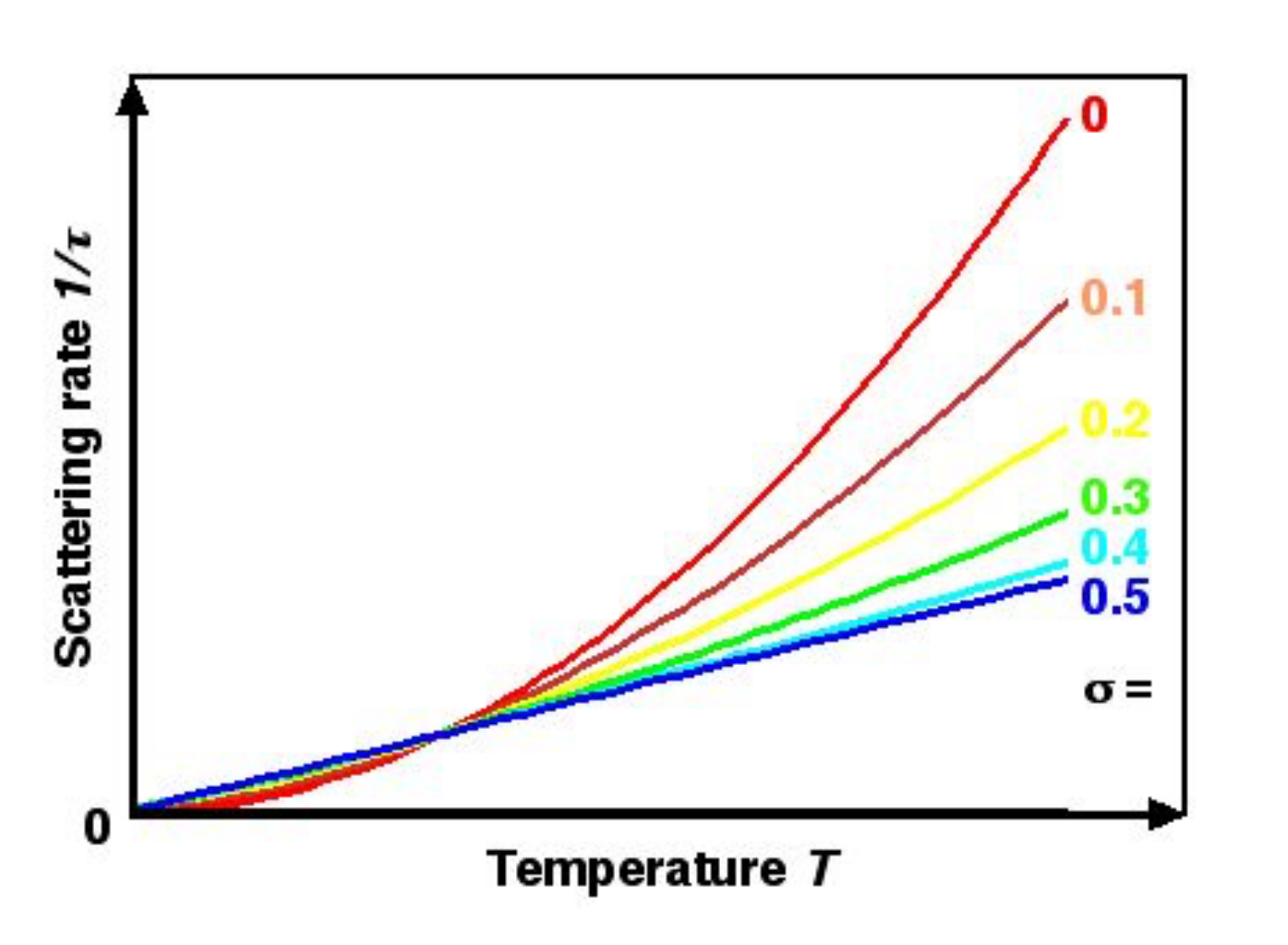}  \footnotesize 

\noindent FIG. 1. Temperature dependence of the scattering rate $1/\tau$ on the relative strength, $\sigma$, of superlattice scattering.  \normalsize

\noindent  Inserted into the Drude formula, this gives
\begin{equation}
  \rho_{e,e} = \frac{m^*}{ne^2\tau}   \approx \frac{m^*}{n} \frac{k_B}{e^2\hbar}\; T\;.
\end{equation}



\begin{thebibliography}{4}

\bibitem{1} L. D. Landau and I. Ya. Pomeranchuk, JETP \textbf{7}, 379 (1937).

\bibitem{2}
I. M. Lifshitz and M. I. Kaganov, Sov. Phys. Uspeki \textbf{8}, 805 (1966).

\bibitem{3} M. Bucher, ``Atomic configuration in cuprates at the closing of the pseudogap,'' arXiv:2110.05931
\bibitem{4} A. Legros, S. Benhabib, W. Tabis, F. Lalibert\'{e}, M. Dion, M. Lizaire, B. Vignolle, D. Vignolles, H. Raffy, Z. Z. Li, P. Auban-Senzier, N. Doiron-Leyraud, P. Fournier, D. Colson, L. Taillefer, and C. Proust, Nat. Phys. \textbf{15}, 142 (2019).

\end{thebibliography}
\end{document}